\documentclass[12pt]{article}%
\usepackage{amsmath}
\usepackage{amsfonts}
\usepackage{amssymb}
\usepackage{graphicx}
\usepackage{xcolor}
\usepackage{float}
\usepackage{color}
\usepackage{geometry}
\usepackage{natbib}%
\providecommand{\U}[1]{\protect\rule{.1in}{.1in}}

\numberwithin{equation}{section}

\numberwithin{equation}{section}

\newcommand{\be}{\begin{equation}}
\newcommand{\ee}{\end{equation}}

\newcommand{\bq}{\begin{eqnarray}}
\newcommand{\eq}{\end{eqnarray}}

\setcitestyle{authoryear,round}
\begin{document}
\title{Joint Liability Model with Adaptation to Climate Change \footnote{*We thank  
Marcel Oestreich, Amr ElAlfy, Chengguo Weng, and Johnny Li, whose constructive comments and suggestions have
improved both the substance and presentation of the paper. The usual disclaimer applies.
We also thank the organizers and the participants at the 4th Waterloo Student Conference in Statistics, Actuarial Science and Finance in Waterloo, Ontario, October 27 - 28, 2023.}}
\author{Jiayue Zhang\thanks{Department of Statistics \& Actuarial Science, University of Waterloo (j857zhan@uwaterloo.ca)}
\and Ken Seng Tan\thanks{Division of Banking and Finance, Nanyang Technological University (kenseng.tan@ntu.edu.sg)} 
\and Tony S. Wirjanto \thanks{Department of Statistics \& Actuarial Science, University of Waterloo (twirjanto@uwaterloo.ca)} \thanks{School of Accounting \& Finance, University of Waterloo (twirjanto@uwaterloo.ca)}
\and 
Lysa Porth \thanks{Gordon S. Lang School of Business and Economics, University of Guelph (lporth@uoguelph.ca)}
}
\date{{\small May 2025 }}
\maketitle
\begin{abstract}
\noindent This paper extends the application of ESG score assessment methodologies from large corporations to individual farmers' production, within the context of climate change. Our proposal involves the integration of crucial agricultural sustainability variables into conventional personal credit evaluation frameworks, culminating in the formulation of a holistic sustainable credit rating referred to as the Environmental, Social, Economics (ESE) score. This ESE score is integrated into theoretical joint liability models, to gain valuable insights into optimal group sizes and individual-ESE score relationships. Additionally, we adopt a mean-variance utility function for farmers to effectively capture the risk associated with anticipated profits. Through a set of simulation exercises, the paper investigates the implications of incorporating ESE scores into credit evaluation systems, offering a nuanced comprehension of the repercussions under various climatic conditions.



\end{abstract}

\makeatletter

\makeatother
\baselineskip 18pt

\newpage

\section{Introduction}

In recent years, the importance of incorporating sustainability factors into financial decision-making has gained significant traction across various sectors of the economy, particularly within agriculture and finance. While Environmental, Social, and Governance (ESG) metrics have become a cornerstone for evaluating corporate sustainability, a notable gap persists in assessing individual borrowers—particularly farmers seeking credit to finance their agricultural operations. To address this gap, this paper introduces a novel concept: a scoring system coined as Environmental, Sustainability, and Economics (ESE). The ESE score adapts the ESG framework to the agricultural sector, emphasizing economic considerations alongside environmental and social aspects as critical components of sustainable farming practices. This approach reflects the unique realities of individual farmers, where governance metrics are less applicable, and financial sustainability is paramount. 

The ESE scoring system is designed to address the needs of multiple stakeholders who stand to benefit from such a framework. Policymakers and regulators can utilize ESE scores to design and implement targeted subsidies or tax incentives that promote sustainable farming practices. Financial institutions, such as banks and insurers, could leverage ESE metrics to assess the creditworthiness and risk profiles of individual farmers, while aligning their lending portfolios with sustainability goals. Similarly, agribusiness corporations can use ESE scores to evaluate their suppliers’ sustainability performance, ensuring greater transparency and alignment with their ESG-related commitments. Most importantly, farmers and cooperatives can benefit directly from the ESE scoring system, as it provides a structured framework to benchmark their practices, improve access to sustainable financing, and signal their commitment to sustainability to potential partners and stakeholders.

Although the ESE score draws inspiration from the ESG framework, there are both shared characteristics and critical differences between the two concepts. Both ESE and ESG scores emphasize sustainability and evaluate entities based on environmental and social metrics, such as resource use efficiency, emissions reduction, and biodiversity impacts. However, while ESG scores are primarily designed for corporations, investors, and financial markets, ESE scores are tailored specifically for the agricultural sector, focusing on individual farmers. A key distinction lies in the explicit inclusion of economic metrics in the ESE score, addressing the financial realities of farming—such as profitability and cost-efficiency—which are not standalone components in ESG. Additionally, ESG scores encompass broader social and governance considerations, such as diversity, labor practices, and organizational governance. In contrast, the ESE score narrows its scope to factors directly relevant to sustainable agriculture and individual creditworthiness. This granularity makes ESE scores a more specialized tool for supporting agricultural stakeholders while complementing broader ESG frameworks.

In this paper, we also propose a novel link between individual ESE scores and a joint liability model, providing lenders and borrowers (where lenders are insurance companies and borrowers are farmers) with optimal scores and corresponding optimal group sizes that maximize the total utility of a group of individual borrowers. We begin our analysis with the simplest case of two borrowers in a group and later generalize it to a group of $n$ individual borrowers. Incorporating the individual ESE score mechanism into the joint liability model offers benefits to both lenders and borrowers. For lenders, this mechanism provides a better understanding of the borrower’s risk profile, enabling informed lending decisions and ensuring more sustainable joint liability contracts. Additionally, it offers insights into optimal group size for lending purposes, enhancing the efficiency of group-based borrowing. For borrowers, this mechanism offers a clearer picture of their ranking relative to peers, allowing them to identify areas for improvement and enhance their individual ESE scores. It also helps borrowers determine the ideal number of peers needed to form a group for borrowing purposes, improving group success rates and access to credit. Therefore, integrating the individual ESE score mechanism into credit evaluation processes can result in better-informed lending decisions, improved credit access, and enhanced sustainable development outcomes.

The remaining parts of this paper are organized as follows. We list the steps and potential metrics for building an individual-ESE score mechanism in Section \ref{ch3.1}. In Section \ref{ch3.2}, we introduce the set-up and assumptions of the model and derive a set of results from the models consisting of a group size of two and a group size of $n$. In Section \ref{ch3.3}, we refine the utility functional form of the model to that of a mean-variance one, which incorporates both the expected value of return and risk simultaneously in the face of climate change. As the resulting model does not have cleanly closed-form analytical expressions, a set of simulation exercises is provided in this paper based on a set of climate scenarios. Section \ref{ch3.4} concludes this paper.

\section{The Mechanism of Individual-ESE Scores}
\label{ch3.1}

In order to maintain accountability for sustainable agriculture commitments and provide valuable insights for policymaking, conducting regular and transparent assessments is of paramount importance. However, there are a number of variations in the definitions of sustainable agriculture, and also a limited number of quantitative assessments of agricultural sustainability that are currently available, according to \cite{agovino2018policy}. Although some experts view sustainable agriculture as a collection of management strategies, others see it as an ideology or a specific set of goals. Despite this, there is an increasing consensus that sustainable agriculture should be evaluated based on its environmental, economic, and social impacts, as emphasized by \cite{ZHANG20211262}. Several frameworks and indicators have been developed. Existing sustainable agriculture indicators mainly assess the environmental impacts of agriculture production, and many have limited data availability. Although various organizations have called for monitoring agriculture globally, actual datasets to enable trend assessments are still currently lacking. The absence of a consistent quantification of agricultural sustainability across multiple dimensions impedes the successful identification of undesirable trade-offs and the development of win-win solutions. Given these challenges, relying on literature-based metrics is a reasonable starting point in developing a sustainable agricultural credit scoring system. Literature-based metrics are metrics that have been previously used and validated in other studies or publications and are therefore based on a larger sample size and a more diverse set of variables than any single data collection effort. Having literature-based metrics about agricultural credit scoring systems allows us to apply them when we obtain real data or use them to create a scoring system by using synthetic data. By compiling these synthetic metrics and adjusting them to fit the specific needs of the agricultural sector, we can create a credit scoring system that is more comprehensive and accurate than any single data collection effort.

The ultimate objective of this research is to eventually establish a comprehensive and sustainable agricultural credit scoring system at the individual level, which encompasses three critical aspects: Economic, Social, and Environmental. This system will be called an individual-ESE score. While traditional credit scoring systems focus mainly on borrowers' economic and financial status, our system will take into account a broader range of factors that contribute to sustainability. We recognize the importance of economic metrics and acknowledge that they are crucial to the classical credit scoring system. Therefore, we will use some of the commonly used economic metrics in the related literature as a foundation for our economic metrics. Presented below are synopses of selected academic papers that have conducted research on the agricultural credit scoring system. (1) \cite{romer2018can} investigated the use of weather data in agricultural credit scoring for microfinance institutions, finding that simple correlation analysis between weather events and loan repayment is insufficient to forecast future repayment behavior. (2) \cite{turvey1991credit} reviewed the use of credit scoring for agricultural loans and its application to agricultural finance, identifying the importance of accurate data and the need for specialized credit scoring models for agricultural lending. (3) \cite{lufburrow1984credit} developed a credit scoring model for farm loan pricing, utilizing statistical methods to determine the significant factors affecting loan repayment and default rates. (4) \cite{limsombunchai2005analysis} analyzed credit scoring for agricultural loans in Thailand, identifying the factors that affect credit scores and their impact on loan approval and interest rates. (5) \cite{bumacov2014use} examined the use of credit scoring in microfinance institutions, reporting that it can improve the outreach and efficiency of microfinance operations, but also highlighting the need for effective risk management strategies. (6) \cite{mutamuliza2021determinants} investigated the determinants of smallholder farmers' participation in microfinance markets in Rwanda, identifying factors such as gender, education, and access to information that can affect farmers' ability to access credit. We provide a summary of the key metrics employed in their credit scoring models as the foundations of economic metrics in Table \ref{tab31}.

\cite{ZHANG20211262} presents a variety of useful metrics for the environmental and social pillars. With additional information provided in \cite{fuglie2016g20} and \cite{movilla2021toward}, we have compiled a comprehensive list of potential metrics for these pillars, along with detailed descriptions and methods of measurement. The metrics are shown in Table \ref{tab32}.

\begin{table}[H]
\begin{tabular}{|p{4cm}|p{7cm}|p{4cm}|}
\hline
\textbf{Metric} & \textbf{Description} & \textbf{Used as a Metric in the Scoring System} \\
\hline
\hline
Age & Age of the applicant in years & (1), (4) \\
\hline
Gender & Male, Female, Others & (1), (6) \\
\hline
Education & Education level of the applicant & (6) \\
\hline
Collateral & Collateral in thousands & (1), (3), (4)  \\
\hline
Assets & Assets in thousands & (1), (2), (3), (4), (5) \\
\hline
Debt & Debt to other banks in thousands & (1), (2), (3), (4), (5)   \\
\hline
Deposit & Deposit in the bank account in thousands & (1), (2), (5)  \\
\hline
Income & Monthly business and household income in thousands & (1), (2), (3), (4), (6)   \\
\hline
Marital status & Single, Married, Divorced, Others & (1) \\
\hline
No. family members & Number of family members & (1)\\
\hline
Loan Size & Loan amount in thousands & (1), (4), (6) \\
\hline
Loan duration & Loan Duration in Months & (4) \\
\hline
Repayment ability & Projected net cash flow plus projected inventory divided by a total line of credit & (1), (2), (3)  \\
\hline
Repayment history & The average of loan principal repaid divided by principal due over the past three years & (1), (3) \\
\hline
Tenure & The ratio of acres owned and farmed to total acres farmed & (3), (6) \\
\hline
Resident & 1 if applicant is a resident; 0 otherwise & (1), \\
\hline
Working experience & Working experience in current profession in months & (1),  \\
\hline
Region and Country & Region and Country of the Agricultural Business & (1), (2), (4), (6)  \\
\hline
Distance & Distance to the microfinance institution in km & (6)\\
\hline
\end{tabular}
\centering
\caption{A List of Economic (E) Metrics Applied in Literature}
\label{tab31}
\end{table}

Developing a comprehensive agricultural credit scoring system requires careful consideration of the various metrics that influence creditworthiness. Below is a scheme for generating a composite score based on multiple metrics, including how to assign weights to each metric. We aim at eventually developing a comprehensive ESE scoring system by either obtaining real data or generating synthetic data according to this scheme in the future.

\begin{enumerate}
    \item Identify and gather relevant metrics: Begin by identifying the metrics that are most critical to assessing creditworthiness in agricultural lending, such as past repayment history, crop yields, farm management practices, and market conditions.
    \item Determine the relative importance of each metric: Assign weights to each metric based on its perceived importance in predicting creditworthiness. This can be done through a combination of expert judgment and statistical analysis of historical data. Weights may also be assigned based on the level of risk exposure associated with each ESE factor, with higher weights assigned to metrics that pose a greater risk to a farmer's performance. Notable, the overall Thomson Reuters ESG Score is calculated by applying an automated, factual logic that determines the weight of each category. Each category consists of a different number of measures. The count of measures per category determines the weight of the respective category. 
    \item Normalize the metrics to ensure that each has the same scale, range, and units of measurement. This can be achieved through methods such as standardization or normalization.
    \item Once the weights and normalized values of each metric have been determined, calculate the composite score by multiplying the weight of each metric by its normalized value and summing the products.
    \item Validate the composite score by comparing it to actual default rates or repayment history. Refine the weights and metrics as needed based on the validation results.
    \item Continuously monitor and update the credit scoring system by incorporating new data and adjusting weights and metrics as needed to improve the accuracy and predictive power of the model.
    \item By following the above steps, a comprehensive agricultural credit scoring system can be developed that considers multiple metrics and assigns weights based on their relative importance in predicting creditworthiness.
\end{enumerate}

\begin{table}[H]
\begin{tabular}{|p{1.2cm}|p{3cm}|p{11.5cm}|}
\hline
\textbf{Sector} & \textbf{Metric} & \textbf{How to Measure}\\
\hline
\hline
E & Soil health & Measuring the quality and health of the soil, including factors such as nutrient levels, organic matter content, and erosion rates. Soil testing kits could be used to measure.\\
\hline
E & Water Consumption & Modelled evapotranspiration using climate data (adjusted for crop type and water availability). Water usage tracking software could be used as a tool to measure. \\
\hline
E & Water Quality & N \& P concentration (measured in
stream, or modeled using land cover for landscapes) \\
\hline
E & Habitat Conversion & Remotely sensed land use (\% of study area covered by natural habitat)\\
\hline
E & Habitat Composition & Remotely sensed land use (no. of land cover classes, indicating habitat diversity) \\
\hline
E & Habitat Connectivity & Remotely sensed land use (calculated connectivity score) \\
\hline
E & Species Richness & Field samples (richness relative to reference natural landscape) \\
\hline
E & Species Abundance & Field samples (abundance relative to reference natural landscape, i.e. the most
undisturbed nearby similar habitat) \\
\hline
E & Yield / Area & Harvested mass of crop per unit area over a 5-year rolling average, perhaps using geometric mean to penalize variation in yield. \\
\hline
E & Use of pesticides and chemicals & Liters of pesticide or chemicals used per hectare\\
\hline
E & Carbon Footprint & The carbon footprint for different crop types are shown in \cite{sah2018carbon}. Measuring the amount of greenhouse gas emissions generated from the farmer's activities, including fertilizer use, transportation, and energy use. \\
\hline
S & Health \& safety issues & Whether the applicant ever had safety issues while participating in agricultural activities (0 if no, 1 if yes). \\
\hline
S & Insuring health & Whether the applicant has personal accident insurance or commercial insurance (0 if no, 1 if yes).\\
\hline
S & Labor practices & Measuring the fairness and safety of labor practices on the farm, including minimum wage compliance, safety regulations, and worker satisfaction. \\
\hline
S & Community engagement & Measuring the level of engagement and support the farmer provides to the local community, including economic benefits, charitable contributions, and community outreach. \\
\hline
S & Ethical business practices & Measuring the farmer's adherence to ethical business practices, including transparency, honesty, and accountability.\\
\hline
S & Compliance & Measuring the farmer's compliance with local and national regulations, including environmental, labor, and health and safety regulations.\\
\hline
\end{tabular}
\centering
\caption{A list of Environmental (E) and Social (S) Metrics}
\label{tab32}
\end{table}

\section{The Framework for an Individual ESE-Joint Liability Model}
\label{ch3.2}

We are interested in examining a microcredit lending scenario in which a benevolent lender extends loans to a group of self-selected micro-entrepreneur borrowers who are jointly liable for repayment of the loan. Although each member of the group is given an individual loan, the group is collectively responsible for repaying the loans, and if the total loan is repaid, the entire group can qualify for a subsequent loan. The loans are then invested in a number of projects, which have an equal chance of success and can be either disjoint from, or correlated with, each other. It is assumed that the group members know the realized output of each project, while the lender does not. Both the lender and the borrowers have to make decisions in this setup. The lender has to decide on the optimal contract structure, the degree of joint liability, and the level of punishment in the event of a strategic default, while the borrowers have to decide whether to repay or default, the loan. 

There are two prior studies which have inspired us in particular to write this paper. First \cite{rezaei2017optimal} show that a group size can affect lending efficiency. Larger groups are more reliable in repaying loans for riskier projects, but the group size cannot become too large. It is shown that joint liability has a positive effect on borrowers' repayment amount and welfare compared to individual liability, provided that group members impose strong punishment strategies against any strategically defaulting member. In addition, \cite{rezaei2017optimal} also examine the possibility of relaxing some of the simplifying assumptions made in the basic model of joint liability. 
Second, in the context of climate change,  \cite{castaing2021joint} investigates the impact of risk--sharing networks on the adoption of weather shock management strategies in Burkina Faso; particularly, she studies how the principle of forced solidarity may reduce efforts to adopt techniques that mitigate exposure to climate change. The study finds that a system based on mutual assistance between farmers may reduce their incentives to adopt risk--mitigating strategies, both incrementally and transformationally. The study highlights the need for complementary mechanisms, such as insurance, to help farmers reduce their vulnerability to climate change.

Following the aforementioned two studies, this paper pursues two primary objectives. The first one is to determine an optimal group size that maximizes the benefit for the borrowers, while ensuring that the lender breaks even. A larger group size can have a positive effect on the expected repayment amount, as more individuals are liable for repaying defaulted payments, thus ensuring a higher rate of repayment. However, a larger group size can also pose a threat to repaying members, as they must repay all defaulting peers' repayments, when everyone else in the group defaults on repayment. Therefore this paper proposes a method to determine the optimal group size that enables joint liability to achieve its maximum loan ceiling. 
A second objective of this paper is to derive an optimal individual--ESE score. The importance of such a scoring system for both lenders and borrowers cannot be overstated. For lending companies, having access to a specific scoring mechanism for individual borrowers can provide a better understanding of their creditworthiness and repayment prospects. This information is essential for lending companies in determining the risk associated with lending to a particular borrower. Additionally, the optimal group size of borrowers can be determined using the scoring system, which enables the lending company to mitigate risk by lending to a well-diversified portfolio. For borrowers, having a scoring system can equip them with a clearer picture of their ranking and eligibility for loans. This information is critical for farmers as they search for an ideal group for borrowing purposes since this can help them determine the number of peers needed in a group to secure a loan and, in the process, improve their chances of obtaining funding. Therefore, both lenders and borrowers can benefit from the implementation of such a comprehensive scoring system.

\subsection{Model Assumptions}

Below we state a number of assumptions underlying the proposed model. 

\noindent {\bf ASSUMPTION 1}. The borrowers (who are the individual farmers) produce an agricultural product. For each borrower, the probability of high production (i.e., a successful project) is the same as given by $e$ and the probability of low production (i.e., a failed project) is given by $1-e$ respectively.

\noindent {\bf ASSUMPTION 2}. A higher individual--ESE score means that the farmer has a stronger ability to succeed in their project due to a high-quality comprehensive evaluation. For simplicity, the probability of high yield is assumed to be linearly related to the borrower's ESE score; i.e., we assume that there is a linear relationship between $e$ and $ESE$: $e=k \cdot E S E+b$. The parameter $k$ represents the weight or strength of the impact of the borrower's ESE score on the probability of high yield. A higher value of $k$ indicates that a higher ESE score leads to a more significant positive impact on the probability of success. This aligns well with the idea that a borrower's positive ESE practices contribute to a project's success. In other words, higher ESE scores indicate a better alignment with responsibly sustainable practices. The constant $b$ represents a baseline/constant contribution of factors beyond an individual's control, such as climatic and geographical conditions. Even if a borrower has a lower ESE score, this constant term acknowledges that external factors can still influence the project's success. Therefore it is essential to account for such factors to avoid unfairly penalizing borrowers for circumstances beyond their own control.

The selection of the linear expression $e = k \cdot ESE + b$ is motivated by both the characteristics of the variables involved and its practical applicability. Notably, the non-negativity of $k$, $b$, and $ESE$ ensures that the resulting probability, denoted as $e$, maintains its intrinsic non-negative nature. As $ESE$ encompasses a continuous range from 0 to 100, it inherently aligns with this non-negative property. It is important to stress that the technical requirement for $e$, being a probability, to lie between 0 and 1, is conveniently met by setting $b = 0$ and $k = \frac{1}{100}$. In addition, this specification also accommodates real-world situations by allowing for a non-zero baseline influence ($b > 0$), often linked to exogenous factors. To this end, a modest value such as $b = 0.3$ can be accommodated without violating the aforementioned probability constraint. Thus this adjustment necessitates setting $k = \frac{0.7}{100}$, thereby guaranteeing that $e$ remains within the bounds of a probability as required. While, at the same time, it also acknowledges that $e$ cannot span the interval from 0 to 0.3; this constraint is justified by underscoring that $e$ is a representation of a probability of a high yield, which is inherently a positive entity in practical settings. In summary, the rationale underlying the proposed equation effectively marries mathematical coherence with nuanced demands from real-world applications.

At first sight, the use of a sigmoid function to model $e$ in the model as $e = \frac{1}{1 + \exp^{-ESE}}$ or $e = \frac{1}{1 + \exp^{-(k \cdot ESE + b)}}$ appears to be an ideal choice. However, upon a closer examination, a critical observation related specifically to our study emerges; i.e., under the assumption that $k$, $b$, and $ESE$ are non-negative, the use of the sigmoid function would result in a constraint where $e$ precludes the values below $1/2$. This deviation from the alignment with realistic scenarios and the entire spectrum of probability values was deemed to be undesirable in the context of our study. Thus, while the use of the sigmoid function to model $e$ may possess certain merits in other contexts, its inherent restriction on precluding the probability values above $1/2$ in our model is at odds with the overarching goal of accurately representing the probability of high yield across its entire spectrum. As a result, a decision was made to retain the linear equation $e = k \cdot ESE + b$, which both adheres to the established criteria and provides the desired flexibility to represent a plausible range of probability values.
    
\noindent {\bf ASSUMPTION 3}. The amount of a loan secured by each borrower is denoted as $L$, while the amount for this borrower to be repaid is denoted as $w$. The unit selling price of the product is denoted as $p$, and the total income of a high production is denoted as $p \overline{Y}$, while the total income of a low production is denoted as $p \underline{Y}$.

\noindent {\bf ASSUMPTION 4}. High-yield borrowers repay their loans and low-yield borrowers pay back all of the income they earn. As long as this group repays the total loans, this group can continue to secure subsequent loans. In the event that all borrowers are unsuccessful in their projects, the group will not be eligible for any future loans. This may result in a negative credit history for all of the members in the group, which could lead to further restrictions placed on them, such as being unable to use public transportation or purchasing high-cost items. Therefore, if all borrowers fail in their projects, they may have to use all of their earnings to avoid receiving negative credit reports, even though their total earnings will be less than the total repayments.

\noindent {\bf ASSUMPTION 5}. To increase the successful probability $e$, the borrower incurs additional costs, such as purchasing green pesticides, advanced tools, etc. A convex disutility function, denoted by $C(e)$, is included in the specification of the total utility function in the proposed model.

\subsection{Joint Liability Contracts with a Group Size of Two}

Under a joint liability agreement with a group size of two (`Farmer A' and `Farmer B'), from the view of `Farmer A', his/her ex-ante expected profit $\pi_{i}$ can be written as 

\begin{equation}
\pi_{i}=e^2[p \overline{Y}-w]+e(1-e)[p \overline{Y}+p \underline{Y}-2 w]-C(e)
\label{equ31}
\end{equation}

We emphasize that, in this model, both farmers have a chance of achieving a successful harvest $\overline{Y}$ with a probability of $e^2$, which gives them a profit of $p \overline{Y}-w$. The probability of one farmer defaulting is given by $e(1-e)$, which means that farmer $i$ would receive his/her own surplus from a successful harvest minus the other farmer's deficit, $p\overline{Y}+p\underline{Y}-2w$. If both farmers fail in their respective projects at the same time, they will pay all of their incomes to the lenders, which means that their profit will be zero. The detailed profit distributions are summarized conveniently in Table \ref{tab332}. The probabilities and profit outcomes are calculated entirely from the perspective of Farmer A, reflecting the cooperative repayment structure under the assumption that high-yield borrowers (successful farmers) repay their loans in full, while low-yield borrowers (unsuccessful farmers) repay only the income they earn ($p \underline{Y}$). When Farmer A succeeds ($p \overline{Y}$) and Farmer B fails, Farmer A must not only repay their own loan obligation $(w)$ but also cover the shortfall left by Farmer B, who repays only $p \underline{Y}$. This results in Farmer A's profit being $p \bar{Y}-w-(w-p \underline{Y})=p \overline{Y}+p \underline{Y}-$ $2 w$. Conversely, when Farmer A fails and Farmer B succeeds, Farmer A's repayment is limited to their income $(p \underline{Y})$, with Farmer B's success ensuring the cooperative repayment obligation is met, leading to a profit for Farmer A of $p \underline{Y}-p \underline{Y}=0$. The asymmetry in profits arises from this risk-sharing mechanism, where successful farmers are expected to shoulder additional responsibility to maintain the system's financial integrity, while unsuccessful farmers contribute as much as their income allows them to do so, shielding them from excessive financial burdens.

\begin{table}[H]
\begin{tabular}{|p{8.7cm}|p{2.7cm}|p{3.5cm}|}
\hline
\textbf{Case Description} & \textbf{Probability} & \textbf{Profit=Earning-Payment}\\
\hline
\hline
`Farmer A' succeeds, `Farmer B' succeeds too. & $e^2$ & $p \overline{Y}-w$ \\
\hline
`Farmer A' succeeds, `Farmer B' fails. & $e(1-e)$ & $p \overline{Y}+p\underline{Y}-2w$ \\
\hline
`Farmer A' fails, `Farmer B' succeeds. & $(1-e)e$ & $p \underline{Y}-p\underline{Y}=0$ \\
\hline
`Farmer A' fails, `Farmer B' fails. & $(1-e)^2$ & $p \underline{Y}-p\underline{Y}=0$ \\
\hline
\end{tabular}
\centering
\caption{Profit Distributions}
\label{tab332}
\end{table}

Before setting the total objective (utility) function for the whole group in the model, we first list a number of exogenous constraints which we introduce to the optimization program.

The constraints used here are motivated by the commonly accepted assumptions in the literature on joint liability models. For example, \cite{rezaei2017optimal} introduced similar constraints in their analysis of the optimal group size in the joint liability contracts to ensure the feasibility and sustainability of the contract. These constraints are based on expected returns and the repayment strategies of group members. While our study adopts a similar structure for these constraints, the detailed settings of the expected profit and the underlying model assumptions differ from their study. Moreover, our model explicitly incorporates sustainability features, addressing dynamic adjustments and long-term stability of joint liability contracts under varying economic conditions and group compositions.

\noindent {\bf CONSTRAINT 1}. If the group wants a subsequent loan from the lender, the group's total repayment amount must be affordable even in the worst--case scenario when everyone else's project has failed except for him/hers:
    \begin{equation}
    \text{Constraint 1: } 2 w \leqslant p \overline{Y}+p \underline{Y}
    \end{equation}

\noindent {\bf CONSTRAINT  2}. Strategic default in a joint liability contract occurs when one party (or more parties) intentionally chooses (or choose) to default on his/her obligation (or their obligations), even though he/she has (or they have) the ability to pay back the loan. This is often done in order to gain a strategic advantage, such as to force the other parties to pay a greater share of the debt or renegotiate the terms of the contract. To avoid a strategic default for each successful borrower, the payoff of strategically defaulting cannot be larger than the payoff of repaying and being refinanced.

Assuming that the other borrower also succeeds in his/her project, this borrower will only be responsible for his/her own payment. In this situation, the earnings plus the discounted cost of the next loan would at least be equal to the earnings of strategically defaulting.
       \begin{equation}
       \text{Constraint 2: } p \overline{Y} \leqslant p \overline{Y}-w+\delta L
       \end{equation}

\noindent {\bf CONSTRAINT 3}. If the other borrower fails in his/her project, then this borrower needs to pay for himself/herself as well as part of the other borrower's repayment amount. Now, the earnings plus the discounted cost of the next loan would also at least equal the earnings of strategically defaulting.
        \begin{equation}
        \text{Constraint 3: } p \overline{Y} \leqslant p \overline{Y}+p \underline{Y}-2 w+\delta L
        \end{equation}

\noindent {\bf CONSTRAINT 4}. The lender must be able to sustain the lending game, or at least break even, over the periods. Thus the expected repayment amount of each borrower has to be at least as large as $L(1+\epsilon)$, where $\epsilon$ is a risk-free rate. It is important to note that given that all constraints have been established under the assumption that the group will successfully repay the total loans to ensure their eligibility for subsequent loan rounds, we have not factored in the case, in which all borrowers within the group experience failure in their projects, on the left-hand side of Constraint 4. 
    \begin{equation} \text{Constraint 4: } w\left[1-(1-e)^{2}\right] \geqslant L(1+\varepsilon)
    \end{equation}

Since $p \underline{Y}-w<0$, it is clear that Constraint 3 is tighter than Constraint 2. In the ensuing analysis, we will only consider Constraint 1, Constraint 3, and Constraint 4. Firstly $\pi_{i}$ and $w$ have a strictly negative relationship as can be observed from Equation (\ref{equ31}). So, to maximize $\pi_i$, $w$ has to be set at a minimum value. From Constraint 4, the minimum value of $w$ should be set at
\begin{equation}
w=\frac{L(1+\varepsilon)}{1-(1-e)^2}
\end{equation}.

Next we substitute $w=\frac{L(1+\varepsilon)}{1-(1-e)^{2}}$ into Constraint 1, and rewrite the equation as

\begin{equation}
\quad 2 \cdot \frac{L(1+\varepsilon)}{1-(1-e)^{2}} \leqslant p \overline{Y}+p \underline{Y}, 
\end{equation}

and the first loan ceiling bound can be expressed in Equation (\ref{equ32}) as

\begin{equation}
    L_1 \leqslant \frac{p \overline{Y}+p \underline{Y}}{2(1+\varepsilon)}\left[1-(1-e)^{2}\right].
    \label{equ32}
\end{equation}

After substituting $w=\frac{L(1+\varepsilon)}{1-(1-e)^{2}}$ into Constraint 3, we obtain

\begin{equation}
    p \overline{Y} \leqslant p \overline{Y}+p \underline{Y}-2 \frac{L(1+\varepsilon)}{1-(1-e)^{2}}+\delta L , 
\end{equation}

\begin{equation}
    \left(\frac{2(1+\varepsilon)}{1-(1-e)^{2}}-\delta\right) L \leqslant P \underline{Y} , 
\end{equation}

This allows us to obtain another bound for the loan ceiling as in Equation (\ref{equ33})

\begin{equation}
    L_2 \leqslant \frac{p \underline{Y}}{\frac{2(1+\varepsilon)}{1-(1-e)^{2}}-\delta}
    \label{equ33}
\end{equation}

Next we substitute $w=\frac{L(1+\varepsilon)}{1-(1-e)^{2}}$ into the objective function (which is Equation (\ref{equ31}) and replace $e=k \cdot E+b$. The ex-ante expected profit function of an individual borrower can now be rewritten as

\begin{equation}
\begin{aligned}
& \max _E e^2(p \overline{Y}-w)+e(1-e)[p \overline{Y}+p \underline{Y}-2 w]-C(e) \\
& =e^2 p \overline{Y}+e(1-e)(p \overline{Y}+p \underline{Y})-\left(2 e-e^2\right) w-C(e) \\
& =e^2 p \overline{Y}+e(1-e)(p \overline{Y}+p \underline{Y})-L(1+\varepsilon)-C(e) \\
& = (k E+b)^2 \cdot p \overline{Y}+(k E+b)(1-k E-b)(p \overline{Y}+p \underline{Y})-L(1+\varepsilon)-C(e)
\label{equ34}
\end{aligned}
\end{equation}

To simplify the problem even further, we set $C(e)=\frac{1}{2} c e^{2}$ (where the cost parameter $c$ is used to scale the total cost) as the disutility function. The disutility function $C(e)=\frac{1}{2} c e^{2}$ is a commonly used function in Economics to represent the cost or disutility associated with effort or work. See e.g. \cite{clark1996satisfaction} and \cite{kahneman1997back}. The function is chosen to be quadratic so that it can accommodate the following relationship: as an effort increases, the cost or disutility of that effort increases at an increasing rate. In the context of an individual agricultural farmer, it is reasonable to use a disutility function to acknowledge that while some effort may be necessary, there is a point at which additional effort becomes more and more unpleasant or burdensome. For instance, a farmer may be willing to put in extra effort to cultivate crops or tend to livestock, but as the required effort increases, the farmer may experience fatigue, stress, and reduced productivity, which can negatively impact their overall well--being. Furthermore, the quadratic form of the disutility function is also supported by empirical evidence on the relationship between effort and well-being. It is reasonable to assume that farmers will experience a phenomenon known as a diminishing marginal utility of effort, which implies that the additional unit of well-being gained by exerting more effort decreases as the level of effort increases.

\subsection{Main Results}

In this subsection, we state a number of key results expressed conveniently in the form of propositions. These propositions provide useful insights into the optimal individual--ESE scores, optimal group size, and loan ceiling constraints.

\noindent \textbf{PROPOSITION 1 (Optimal individual--ESE score)}.
Finding an optimal ESE score for both the lenders and borrowers in an individual ESE--joint liability model is critical because it can help address the challenge of promoting sustainability in agricultural production while ensuring access to credit for small-scale farmers. By incorporating environmental and social factors into credit evaluation systems, it can incentivize farmers to adopt sustainable practices that can lead to positive environmental outcomes and long-term economic benefits. At the same time, lenders can mitigate the risks associated with lending to farmers by incorporating sustainability factors into their credit evaluation process.

To find the optimal value of individual--ESE score, $E$, we take a first-order partial derivative with respect to $E$ of the ex-ante expected payoff function to obtain the following result

\begin{equation}
\begin{aligned}
\frac{\partial \pi}{\partial E} & =2(k E+b) \cdot k(p \overline{Y})+(k-2(k E+b) \cdot k) \cdot[p \overline{Y}-p \underline{Y}]-k \cdot \frac{\partial C(e)}{\partial e} \\
& =k p \overline{Y}-[k-2(k E+b) \cdot k] p \underline{Y}-k c \cdot(k E+b) = 0.
\end{aligned}
\end{equation}

After rearranging terms, we obtain an optimal individual-ESE score (where $0 \leqslant E \leqslant 100$) as 

\begin{equation}
E= \frac{p \overline{Y}+(2 b-1) p \underline{Y}-c b}{c k-2 k p \underline{Y}}
\label{equ35}
\end{equation}

\noindent {\textbf{PROPOSITION 2 (Collapsing two constraints on loan ceiling to a single constraint)}.
The two loan ceiling constraints, Constraint 1 and Constraint 3, originate from analyzing different worst-case scenarios to ensure the affordability and sustainability of the group’s repayment structure. However, most of the previous studies in this area have not explored the connection between these two constraints. Notably, in \cite{rezaei2017optimal}, they found that a feasible range of loan ceilings depends on the borrower's discount rate $\delta$. Nevertheless, it is possible to reduce these two constraints to a single constraint in our model.

Recall the two constraints on the loan
ceiling in Equation (\ref{equ32}) and Equation (\ref{equ33}), labelled respectively as $L_1$ and $L_2$. 
Next, we show that $L_1$ is always larger than $L_2$, which means that we can only use $L_2$ as a final loan ceiling bound. It simplifies the problem and potentially makes it easier to solve the model. In addition, $L_2$ takes $\delta$ into consideration, which may include more information and additional key factors. This result is illustrated in Figure \ref{fig32}. 

\begin{figure}[!htp]
\includegraphics[width=14cm]{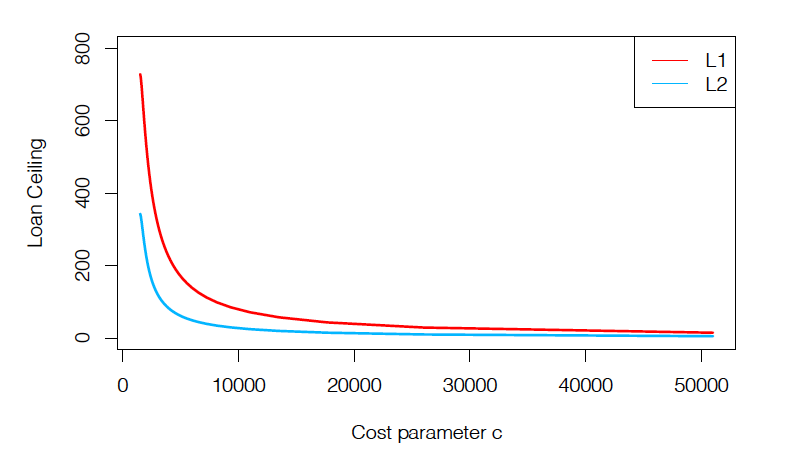}
\centering
\caption{The Relationship between Two Loan Ceiling Constraints}
\label{fig32}
\end{figure}

\textbf{Proof:}

If $L_{1} \leqslant L_{2}$, then

\begin{equation}
    \frac{p \overline{Y}+p \underline{Y}}{2(1+\varepsilon)}\left[1-(1-e)^{2}\right] \leqslant \frac{p \underline{Y}}{\frac{2(1+\varepsilon)}{1-(1-e)^{2}}-\delta}
\end{equation}

After rearranging terms, we obtain

\begin{equation}
    \delta \geqslant \frac{\overline{Y} \cdot 2(1+\varepsilon)}{\left[1-(1-e)^{2}\right](\overline{Y}+\underline{Y})}>1
\end{equation}

From the above results, if $L_{1} \leqslant L_{2}$, the discount factor $\delta$ is always greater than 1, which leads to a contradiction. This leads us to conclude that $L_{1} > L_{2}$. 

\noindent \textbf{PROPOSITION 3 (A higher individual--ESE score leads to a higher loan ceiling)}.
Next, we examine the relationship between the individual--ESE score and the loan ceiling. Intuitively a borrower with a high-quality ESE score should be eligible for a higher loan amount, and therefore the loan ceiling should also be higher.

Since $L_1$ is always greater than $L_2$, we will use $L_2$ to determine the loan ceiling, which defines the maximum value of $L_2$ to be the loan ceiling.

\begin{equation}
\begin{aligned}
    L_{2} & = \frac{p \underline{Y}}{\frac{2(1+\varepsilon)}{1-(1-e)^{2}}-\delta} \\
& =\frac{p \underline{Y}\left[1-(1-e)^{2}\right]}{2(1+\varepsilon)-\delta\left[1-(1-e)^{2}\right]} \\
&=\frac{p \underline{Y}\left(2 e-e^{2}\right)}{2(1+\varepsilon)-\delta\left(2 e-e^{2}\right)}
\end{aligned}
\end{equation}

Then we take a partial derivative of $L_2$ with respect to $E$ to obtain

\begin{equation}
    \begin{aligned}
\frac{\partial L_{2}}{\partial E}&=\frac{\partial L_{2}}{\partial e} \cdot \frac{\partial e}{\partial E} \\
&=\frac{\left.p \underline{Y}(2-2 e) \cdot\left[2(1+\varepsilon)-\delta\left(2 e-e^{2}\right)\right]-p \underline{Y}\left(2 e-e^{2}\right)(-\delta\right)(2-2 e)}{\left[2(1+\varepsilon)-\delta\left(2 e-e^{2}\right)\right] ^2} \times k
\end{aligned}
\end{equation}

It is clear that the denominator is greater than 0. The numerator is also greater than 0, which can be illustrated more clearly by writing out the following expression

\begin{equation}
    \begin{aligned}
\text {Numerator } &=k p \underline{Y}(2-2 e) \times\left[2(1+\varepsilon)-\delta\left(2 e-e^{2}\right)+\delta\left(2 e-e^{2}\right)\right] \\
&=k p \underline{Y}(2-2 e)[2(1+\varepsilon)]>0
\end{aligned}
\end{equation}

A borrower with a higher optimal ESE score, which means that he/she has a better overall performance, will be perceived as less risky by the lender. As a result the borrower will be offered a higher loan ceiling, which is the maximum amount of loan that a lender is willing to provide to a borrower. In turn, if a borrower wants to obtain a larger loan amount, he/she may be motivated to improve his/her ESE performance. 

\subsection{Joint Liability Contracts with a Group Size of $n$}

In this subsection, we present an individual ESE--joint liability model which formalizes the assumption that the group of borrowers is collectively responsible for the entire group loan (i.e., $n w$). The expected utility of a borrower $i$ who adopts a repayment strategy during any period when they receive financing is specified as

\begin{equation}
\begin{aligned}
\max _{E} \text{ } \pi_{i}&=\sum_{k=0}^{n-1}\left(\begin{array}{c}
n-1 \\
k
\end{array}\right) \cdot\left[e^{n-k}(1-e)^{k}\right] \cdot\left[p \overline{Y}-w-\frac{k(w-p \underline{Y})}{n-k}\right]-C(e)\\
&=e(p \overline{Y}-w)-(w-p \underline{Y}) \cdot \sum_{k=0}^{n-1}\left(\begin{array}{c}
n-1 \\
k
\end{array}\right) \cdot\left[e^{n-k}(1-e)^{k}\right] \cdot \frac{k}{n-k}-C(e)\\
&=e(p \overline{Y}-w)-(w-p \underline{Y}) \cdot\left((1-e)-(1-e)^{n}\right) - C(e)\\
&=e p \overline{Y}-w(1-(1-e)^{n})+p \underline{Y}\left((1-e)-(1-e)^{n}\right)-C(e)\\
&=(k E+b) p \overline{Y}-w\left(1-(1-k E-b)^{n}\right)+p \underline{Y}\left[(1-k E-b)-(1-k E-b)^{n}\right] - C(e)
\label{equ36}
\end{aligned} 
\end{equation}

Constraints similar to those in the case of a group size of two are as follows (where g stands for `general').

\noindent {\bf CONSTRAINT 1g}. There is at least one borrower who is successful at his/her project and capable of paying the entire group's repayment amount.
    \begin{equation}
    n w \leqslant p \overline{Y}+(n-1) p \underline{Y}
    \end{equation}

\noindent {\bf CONSTRAINT 2g}. Since Constraint 3 is more restrictive than Constraint 2g when the group size is two, we will omit Constraint 2g from further consideration for a group size of \( n \).

\noindent {\bf CONSTRAINT 3g}. For each successful borrower, the payoff of strategically defaulting cannot be larger than the payoff of repaying and being refinanced.
    \begin{equation}
    p \overline{Y} \leqslant p \overline{Y}+(n-1) p \underline{Y}-n w+\delta L
    \end{equation}

\noindent {\bf CONSTRAINT 4g}. The lender must be able to sustain the lending game, or at least break even, over the periods. Thus the expected repayment amount of each borrower has to be at least as large as $L(1+\varepsilon)$, where
$\varepsilon$ is the risk-free rate. As in the two-borrowers case, it is important to note that, given that all of the constraints have been established under the presumption that the group will successfully repay the total loans to ensure their eligibility for subsequent loan rounds, we have not factored in the case, where all borrowers within the group experience failure, on the left-hand side of this constraint.
    \begin{equation}
    w\left[1-(1-e)^n\right] \geqslant L(1+\varepsilon)
    \end{equation}
    
Similarly we obtain $w \geqslant \frac{L(1+\varepsilon)}{1-(1-e)^n}$. Thus the value of $w$ should be set as $\frac{L(1+\varepsilon)}{1-(1-e)^n}$ due to the negative relationship between $w$ and $\pi_{i}$.

After substituting $w = \frac{L(1+\varepsilon)}{1-(1-e)^{n}}$ in the objective function (which is Equation (\ref{equ36})), the ex-ante expected payoff function for the borrower $i$ in a size $n$ joint liability group is represented as

\begin{equation}
\max _E \pi_{i}=(k E+b) p \overline{Y}-L(1+\varepsilon)+p \underline{Y}\left[(1-k E-b)-(1-k E-b)^n\right]-C(e)
\end{equation}

\noindent \textbf{PROPOSITION 4 (Optimal individual--ESE score and group size $n$ have a negative relationship)}.
A higher ESE score indicates a lower default risk, which means that the borrower is more creditworthy and can obtain a loan with a smaller group size. On the other hand, each member of the group is jointly responsible for the repayment of the loan, and if one member defaults, the other members will have to repay the outstanding amount. Thus, a low-quality borrower may choose to loan with a larger group to reduce the probability of default and minimize the potential burden on any one individual. Next, we verify this intuition based on the following derivations.

We take a first-order partial derivative of the expected payoff with respect to $E$ to obtain an optimal individual--ESE score for a borrower as below

\begin{equation}
    \frac{\partial \pi}{\partial E} =k p \overline{Y}+p \underline{Y}k\left[n(1-k E-b)^{n-1}-1\right]-k \frac{\partial C(e)}{\partial e}=0,
\end{equation}
where the cost is $C(e)=\frac{1}{2} c e^2$.

Since it is hard to find an explicitly analytical form of $n$ and $E$, we adopt an implicit function derivation below. Let $E(n)$ replace $E$ and take the partial derivative with respect to $n$ on both sides of
$k p \overline{Y}+p \underline{Y}\left[n(1-k E-b)^{n-1}-1\right]-k c(k E+b)=0$. Then we obtain the following result

\begin{equation}
    p \underline{Y}\left\{(1-k E(n)-b)^{n-1}+n \cdot \frac{\partial \left[e^{(n-1) \log (1-k E(n)-b)}\right]}{\partial n}\right\}-c\left[e \frac{\partial E(n)}{\partial n}\right]=0
\end{equation}

where,

\begin{equation}
\begin{aligned}
\frac{\partial\left[e^{(n-1) \log (1-k E(n)-b)}\right]}{\partial n} 
& =e^{(n-1) \log (1-k E(n)-b)} \cdot\left[\frac{(n-1) \cdot(-k) \frac{\partial E(n)}{\partial n}}{1-k \cdot E(n)-b}+\log (1-k E(n)-b)\right] \\
& =(1-k(n)-b)^{n-1} \cdot\left[\frac{(n-1) k}{1-k E(n)-b} \cdot\left(-\frac{\partial E(n)}{\partial n}\right)+\log (1-k E(n)-b)\right]  
\end{aligned}
\end{equation}

After reorganizing the formula, we can express the partial derivative of $E(n)$ with respect to $n$ as

\begin{equation}
\frac{\partial E(n)}{\partial n} = \frac{p \underline{Y}\left\{(1-e)^{n-1}[1+n \cdot \log (1-e)]\right\}}{p \underline{Y}\left[n \cdot(1-e)^{n-1} \cdot \frac{(n-1) k}{1-e}\right]+c e}
\label{equ37}
\end{equation}

It is clear that the denominator in Equation (\ref{equ37}) is greater than 0. When $n \geqslant 1$ and $e \in (0,1)$, then $n \log (1-e)<-1$, and
$(1-e)^{n-1}[1+n \cdot \log (1-e)]<0$. Thus,
the numerator is less than 0. As a result $\frac{\partial E(n)}{\partial n} < 0$, which shows that the optimal individual--ESE score and group size $n$ have a negative relationship.

In the light of the negative correlation between the optimal individual--ESE scores and the group size, it is recommended that insurance companies prioritize lenders with higher ESE scores when the group size is smaller to mitigate potential risks. Borrowers with high scores are typically perceived to exhibit positive ESE performance, which leads to higher expectations for sustained success and repayment probability over the long term. Despite the relatively small group size, lenders may still have heightened confidence in the ability of high-scoring borrowers to demonstrate honesty and success and to exhibit willingness to repay their loans. As a result, lenders may consider reducing group size requirements for high-scoring borrower groups. Conversely, as the group size increases, the requirement for lenders' ESE scores may potentially decrease, as a larger group may possess a greater risk mitigation capacity and an increased likelihood of successfully repaying the group loan. Consequently, lenders may opt to reduce the score requirements in such cases.

\noindent \textbf{PROPOSITION 5 (Optimal individual--ESE score converges to a constant level for a large enough group size)}.
Equation (\ref{equ37}) tells us that $\lim_{n \rightarrow \infty} \frac{\partial E(n)}{\partial n}=0$, which means that the optimal individual--ESE score converges to a constant level when the group size reaches a certain large enough number. To find the limit of the optimal score value, we recall the first--order condition of the ex-ante expected payoff, which is 

\begin{equation}
    \frac{\partial \pi}{\partial E} =k p \overline{Y}+p \underline{Y}k\left[n(1-k E-b)^{n-1}-1\right]-k \frac{\partial C(e)}{\partial e}.
\label{equ38}
\end{equation}

Similarly we set $C(e)=\frac{1}{2} c e^{2}$, then 

\begin{equation}
    \frac{\partial C(e)}{\partial e}=c e=c(k E+b).
\end{equation}

When $n \rightarrow \infty$, $n(1-k E-b)^{n-1} \rightarrow 0$, then Equation (\ref{equ38}) approaches $k p(\overline{Y}-\underline{Y})-k c \cdot(k E+b)$. Let $k p(\overline{Y}-\underline{Y})-k c \cdot(k E+b) = 0$, the limit of the optimal ESE score will be expressed as a constant

\begin{equation}
    E \rightarrow\left[\frac{p(\overline{Y}-\underline{Y})}{c}-b\right] / k.
\end{equation}

To illustrate this proposition, we provide a numerical example, which is shown in Figure \ref{fig33}. We set the unit price $p = 1$, the successful probability $e = k \times E +b = 1/100 \times E + 0$, high yield amount $\overline{Y} = 1000$, low yield amount $\underline{Y} = 500$, and perform a simulation under a different group size $n \in [1, 100]$.

\begin{figure}[!htp]
\includegraphics[width=14cm]{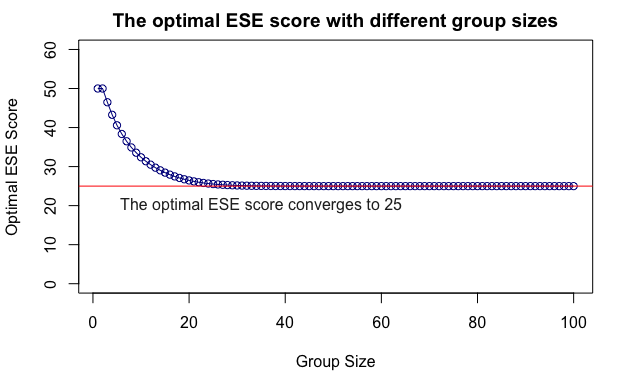}
\centering
\caption{Simulation Results of A Numerical Example}
\label{fig33}
\end{figure}

In situations where the group size is small, insurance companies may seek to attract borrowers with high ESE scores in order to mitigate risk. However, as the group size expands, the marginal benefit of an individual effort on group outcomes diminishes, resulting in a reduction of an individual farmer's optimal effort level. This can create incentives for farmers to contribute less effort, leading to a `free-rider' problem and decreased profits for the group. Therefore the threshold value of the optimal ESE score can serve as a reference point for both the lenders and borrowers, enabling them to avoid a sub-optimal resource allocation and a `free-rider' problem. 

In the analysis above, $n \rightarrow \infty$ serves as a theoretical construct to explore the asymptotic behavior of the optimal ESE score, $E$. However, in practical applications, $n$, representing the group size, is bounded by real-world constraints. Factors such as coordination challenges, logistical limitations, and the effectiveness of group dynamics impose natural upper limits on the group size. From the results of the simulation plot, it is evident that the optimal ESE score converges once $n$ exceeds a certain threshold. Beyond this point, increasing the group size further does not contribute meaningfully to improving the ESE score and may instead exacerbate issues such as the `free-rider' problem. Consequently, the results derived for large $n$ should be interpreted as indicative of the system's behavior in scenarios where $n$ approaches the upper bounds of realistic group sizes. This bounded nature of $n$ ensures the applicability and relevance of the model to real-world contexts.

\section{Joint Liability Contracts under a Mean--Variance Utility Framework}
\label{ch3.3}

When borrowers assess potential projects, they would like to know how likely it is that they will be able to earn a high return or experience a substantial loss. The variability of profits, or the range of possible returns, is an important consideration for these borrowers. If the potential variability of returns is too high or the risk of substantial losses is too large, the borrower may decide not to invest in the venture. To the best of our knowledge, the current literature on joint liability models focuses only on the expected return or expected profit. This means that they concentrate solely on the central point of an individual's profit without taking into account the distribution of potential profits or the variability of these returns. However, focusing only on the expected return can be limiting, as it does not account for the risks. In this paper, we incorporate the risk (proxied by variance) associated with expected profit into our utility function and conduct further analysis of mean-variance utility in the context of the individual ESE--joint liability model. Specifically, the mean-variance utility function for an individual farmer can be incorporated in his/her ex--ante expected profit function in Equation (\ref{equ341}), where $P$ (a random variable) represents the expected profit and $\gamma$ represents a risk aversion parameter of the farmer. 
The choice to limit the group size to two in this section is intended only to serve as a stylized example to illustrate the key concepts of the mean-variance utility framework. Extending the analysis to larger group sizes introduces complexity further, as it requires simultaneously accounting for both the return and risk components across multiple group members. While such an extension would provide a more general analysis, it is beyond the scope of this study and is left for future work.

\begin{align}
\max_{E} \pi_i & = \mathbb{E}(P) - \frac{\gamma}{2}\mathrm{Var}(P) - C(e)\nonumber \\
&=e^2(p \overline{Y}-w)+e(1-e)[p \overline{Y}+p \underline{Y}-2 w]-C(e) \nonumber \\
&-\frac{\gamma}{2}\left\{e^2(p \overline{Y}-w)^2+e(1-e)(p \overline{Y}+p \underline{Y}-2 w)^2\right\} \nonumber \\
&-\frac{\gamma}{2}\left\{-\left[e^2(p \overline{Y}-w)+e(1-e)(p \overline{Y}+p \underline{Y}-2 w)\right]^2\right\} \nonumber\\
&=e[p \overline{Y}+p \underline{Y}-2 w]-e^2(p \underline{Y}-w)-C(e) \nonumber\\
& -\frac{\gamma}{2}\left\{\left(e^2-e^4\right)(p \overline{Y}-w)^2+\left[e-2e^2+2e^3-e^4\right](p \overline{Y}+p \underline{Y}-2 w)^2\right\} \nonumber\\
& +\frac{\gamma}{2}\left\{2\left(e^3-e^4\right)(p \overline{Y}-w)(p \overline{Y}+p \underline{Y}-2 w)\right\} 
\label{equ341}
\end{align}

In order to derive the required optimal individual--ESE score in this setting, we take a first-order partial derivative of the ex-ante profit with respect to the ESE score

\begin{align}
\frac{\partial \pi_i}{\partial E}= & k(p \overline{Y}+p \underline{Y}-2 w)-2 e k(p \underline{Y}-w)-k \frac{\partial C(e)}{\partial e} \nonumber\\
- & \frac{\gamma k}{2}\left\{\left(2 e-4 e^3\right)(p \overline{Y}-w)^2+\left(1-4 e+6 e^2-4 e^3\right)(p \overline{Y}+p \underline{Y}-2 w)^2\right\} \nonumber\\
- & \frac{\gamma k}{2}\{2\left(3 e^2-4 e^3\right)(p \overline{Y}-w)(p \overline{Y}+p \underline{Y}-2 w)\} \\
= & 0\nonumber
\end{align}

Unfortunately, the resulting analytical solution of the ESE score involves a rather complex expression and is difficult to interpret intuitively; so we resort to simulation studies to reveal several insightful patterns.

Recall that the probability of achieving success in a project is given by $e = k \times ESE + b$, wherein the parameters $k$ and $b$ are predetermined. Subsequently, the probability may be disaggregated into two constituent components. Specifically, the product of $k$ and ESE represents a contribution of a personal impact towards a project's success, whereas $b$ represents a constant contribution of exogenous factors beyond individual control. The value of $k$ is utilized to calibrate the impact of ESE on a project's success, while $b$ pertains to the influence of exogenous factors, including climatic and geographical conditions. 

To determine the value of $b$, an Actuaries Climate Index (ACI) for each region can in principle be utilized as a reference.
Briefly, the ACI is based on an analysis of six climate indicators, including high and low temperatures, heavy precipitation, drought, and sea level rise. The index measures changes in these indicators over time and across regions, providing an indication of how the climate is changing in different parts of a region. 

It is commonly understood that favorable climate conditions have a positive effect on crop yields, which, in turn, increases the probability of project success. Conversely, unfavorable climate conditions have an adverse impact on the likelihood of success. Hence, we delve deeper into the impact of varying climate conditions caused by climate change on ESE score requirements. By changing the farmer's degree of risk aversion from 0 to 1 under different cost parameters, a series of simulations are conducted for a high success probability project under favorable climate conditions (where $b = 0.7$ in Figure \ref{fig34}), a neutral probability project under neutral climate conditions (where $b = 0.5$ in Figure \ref{fig35}), and a relatively low probability project under unfavorable climate conditions (where $b = 0.3$ in Figure \ref{fig36}). After observing and comparing these three figures, we conclude our analysis with three findings below. In general, any factor that can increase the success probability of the project and the repayment probability can directly or indirectly reduce the requirements of the ESE score.

By examining Figure \ref{fig34}, Figure \ref{fig35}, and  Figure \ref{fig36}, we notice that, as the probability of success decreases, the ESE score requirement continues to increase if the control cost remains constant; i.e., the probability of a project's success is inversely proportional to the required ESE score. This is because a higher success rate leads to an increased farmers' ability to repay their loans, thus lowering the risk of default for the insurance company. As a result, the insurance company can lower its ESE score requirements when lending to farmers.

From inspecting any one of the three following figures, we see that, as the degree of risk aversion increases, the required ESE score decreases monotonically. This is because an increase in farmers' risk aversion score leads to a more cautious contract behavior, reducing the probability of a strategic default and increasing the probability of repayment. In other words, a higher degree of risk aversion on the part of the farmers implies a lower level of risk for the insurance company, leading to a lower ESE score requirement.

Again, from inspecting any one of the three figures, we observe that an increase in the cost parameter will result in a higher requirement for the ESE score. An increase in the cost parameter raises the production expenditure of farmers and burdens their economic resources, thus reducing their profits. Therefore, to avoid a high default probability associated with high costs, the corresponding ESE score requirement should increase. Hence, as the cost parameter increases, the insurance company should raise its ESE score requirement to reduce the risk of default among farmers.

\begin{figure}[H]
\includegraphics[width=13cm]{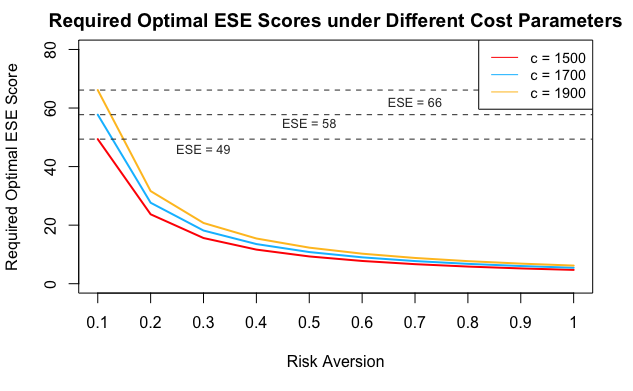}
\centering
\caption{High Successful Probability (Favorable Climate Conditions)}
\label{fig34}
\end{figure}

\begin{figure}[H]
\includegraphics[width=13cm]{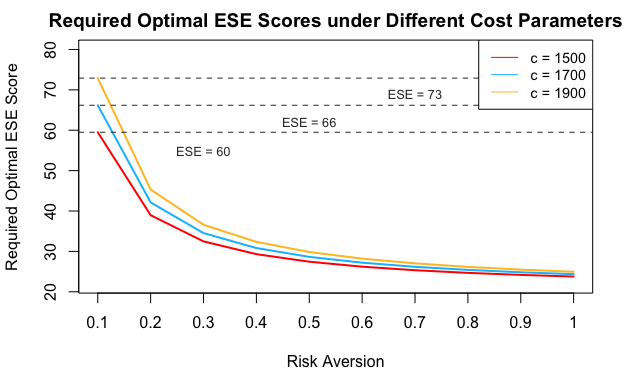}
\centering
\caption{Neutral Successful Probability (Neutral Climate Conditions)}
\label{fig35}
\end{figure}

\begin{figure}[H]
\includegraphics[width=13cm]{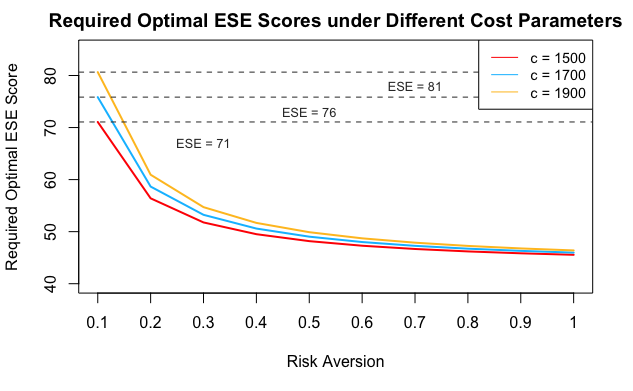}
\centering
\caption{Low Successful Probability (Unfavorable Climate Conditions)}
\label{fig36}
\end{figure}

Finally, we arrive at the question of how changes in crop yields impact the required ESE score. To answer this question, we vary the values of yields and observe the effect on the required ESE score. Specifically we consider two scenarios: a `Low Yield Case' where $\overline{Y} = 1,000$ and $\underline{Y} = 500$, and a `High Yield Case' where $\overline{Y} = 600$ and $\underline{Y} = 300".$ The results of our simulation in Figure \ref{fig37} show that there is a negative correlation between crop yields and the required ESE score, i.e., as crop yields increase, the required ESE score decreases. Conversely, as crop yields decrease, the required ESE score increases. When a borrower has lower yields, this can be taken as an indication of a lower income and a potentially higher risk of default; in other words, a higher requirement for borrowers with lower yields is a reflection of the lender's risk management strategy. By setting stricter requirements for borrowers with lower yields, the lender is attempting to minimize their exposure to financial risk and ensure that the borrowers are able to repay their loans on time and in full. 

\begin{figure}[H]
\includegraphics[width=13cm]{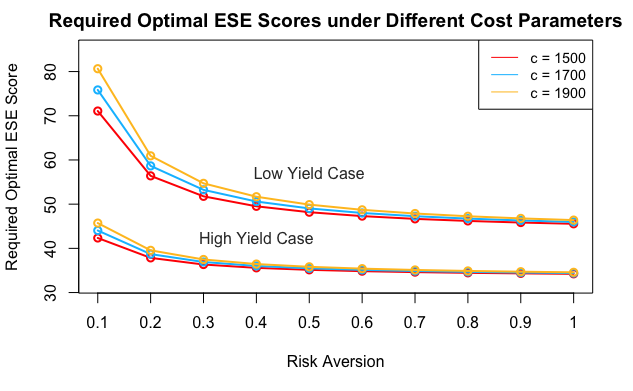}
\centering
\caption{High Yields v.s. Low Yields}
\label{fig37}
\end{figure}

\section{Conclusions}
\label{ch3.4}

In this paper, we have incorporated a novel concept coined as an individual--ESE score into a joint liability model for the purpose of evaluating the sustainability and creditworthiness of individual farmers acting as borrowers in the model in the face of climate change. To achieve this objective we initially considered the expected profit of an individual borrower resulting from the specification of his/her utility function. Our primary goal was to determine the optimal individual--ESE score and to investigate its relationship with key factors such as group size and loan ceiling. By understanding these relationships, we aimed at providing useful insights that benefit both the lending companies and the farmers acting as borrowers. The optimal individual--ESE score serves as a valuable tool for lenders by offering a better understanding of borrowers' risk profiles and informing decisions about appropriate group sizes for lending. For farmers, it can facilitate improved access to credit, thereby fostering greater financial inclusion. Integrating this score into credit evaluation processes not only supports better informed lending decisions but also promotes sustainable development outcomes by aligning financial practices with environmental and social considerations. We listed the main results of this paper conveniently in five propositions. Proposition 1 focused on finding an optimal individual-ESE score. Proposition 2 aimed at collapsing two constraints on the loan ceiling into a single constraint. Proposition 3 demonstrated that a higher individual-ESE score leads to a higher loan ceiling. Proposition 4 revealed that the optimal individual-ESE score and group size $n$ have a negative relationship. Finally, Proposition 5 showed that the optimal individual--ESE score converges to a constant level when the group size reaches a large enough number.

Our study went on to incorporate the risk associated with an expected profit into the utility function of the farmers and analyzed the resulting mean-variance utility framework in the context of the individual ESE--joint liability model. The study additionally disentangled the probability of achieving success in a project into two constituent components, namely the impact of personal factors (represented by $k \times ESE$) and exogenous factors beyond individual farmer's control (represented by $b$), such as climatic and geographical conditions. Our findings reveal that the probability of a project's success is inversely proportional to the required ESE score. This implies that a higher likelihood of a project's success leads to an increased farmers' ability to repay their loans, thereby lowering the risk of default for the insurance company. As a result, the insurance company can lower its ESE score requirements when lending to high-quality farmers. We also found that an increase in farmers' risk aversion score leads to a more cautious contract behavior, reducing the probability of a strategic default and increasing the probability of repayment. Hence, as the risk--aversion score increases, the corresponding ESE score requirement decreases. Finally, an increase in the cost parameter raises the production expenditure of the farmers and places a burden on their economic resources which reduces their profits. Therefore, to avoid a high default probability associated with high costs, the corresponding requirement of an ESE score should be increased.

Lastly, we explored the effects of climate change on the required ESE score in order to ensure a successful project implementation. Our simulation results demonstrated that a higher likelihood of project success, driven by high yields resulting from favorable climate conditions, leads to a lower ESE score requirement in the proposed individual ESE--joint liability model. This suggests that incorporating climate scenarios into the evaluation of creditworthiness can result in better-informed lending decisions and ultimately contribute to more favorable sustainable development outcomes.

As an avenue for future research, this paper paves the way for an exploration of a partially joint liability legal framework that distributes shared responsibility among parties for specific obligations. Unlike a complete joint liability framework, a partially joint liability framework assigns a proportionate debt share to each party. Moreover, the concept of a flexible joint liability introduces aspects of innovation. Participants begin with a collaborative lending similar to a standard joint liability contract but, in strategic default scenarios, opt for a punitive phase rather than immediate maximum penalties. After a defined period ($T$ as previously mentioned), defaulters are reintegrated, which offers a path for forgiveness and adherence to the flexible joint liability contract. The diverse joint liability arrangements highlight the need for a thorough and separate analysis. Future research could explore the application of various joint liability frameworks to enhance understanding and implementation in this area.

Furthermore, to extend the model to appeal to the actuarial domain, it can incorporate key actuarial principles such as risk pooling, premium calculation, and correlated risks. Introducing loss distributions and reinsurance mechanisms would align the model with actuarial practices for assessing extreme losses and managing shared liabilities. The model can also include dynamic economic factors like interest rate and income to capture long-term liability risks, while stochastic optimization techniques could be employed to balance group composition and heterogeneity. Additionally, actuarial approaches for valuing contingent liabilities, setting capital requirements, and integrating ESG factors would make the model more robust and applicable to real-world financial and risk management scenarios. These enhancements would provide practical tools for actuaries to assess, mitigate, and optimize joint liability frameworks effectively.

\newpage

\bibliographystyle{elsarticle-harv}
\bibliography{main}

\end{document}